\documentclass[fleqn,usenatbib]{mnras}

\usepackage{newtxtext,newtxmath}
\usepackage[T1]{fontenc}
\usepackage{ae,aecompl}
\usepackage{hyperref}

\usepackage{graphicx}	% Including figure files
\usepackage{amsmath}	% Advanced maths commands

\usepackage{amssymb}	% Extra maths symbols
\usepackage{xcolor}
\usepackage{hyperref}
\usepackage{orcidlink}
\usepackage{array} 
\usepackage{tabularx}

\newcommand{\orcid}[1]{\href{https://orcid.org/#1}{\textcolor[HTML]{A6CE39}{\aiOrcid}}}

\def\la{\mathrel{\mathpalette\fun <}}
\def\ga{\mathrel{\mathpalette\fun >}}
\def\fun#1#2{\lower3.6pt\vbox{\baselineskip0pt\lineskip.9pt
        \ialign{$\mathsurround=0pt#1\hfill##\hfil$\crcr#2\crcr\sim\crcr}}}

\title[HOD and SAM for galaxy clustering]{Clustering analysis of BOSS-CMASS galaxies with semi-analytical model for galaxy formation and halo occupation distribution}

\author[Z. Zhai et al.]{\parbox{\textwidth}{
Zhongxu Zhai,$^{1,2,3,4}$\thanks{E-mail: zhongxu.zhai@uwaterloo.ca}
Andrew Benson, $^{5}$
Yun Wang, $^{6}$
}\vspace*{4pt}\\
$^{1}$Department of Astronomy, School of Physics and Astronomy, Shanghai Jiao Tong University, Shanghai 200240, China \\
$^{2}$Shanghai Key Laboratory for Particle Physics and Cosmology, Shanghai 200240, China \\
$^{3}$Waterloo Center for Astrophysics, University of Waterloo, Waterloo, ON N2L 3G1, Canada \\
$^{4}$Department of Physics and Astronomy, University of Waterloo, Waterloo, ON N2L 3G1, Canada \\
$^{5}$Carnegie Observatories, 813 Santa Barbara Street, Pasadena, CA 91101, USA \\
$^{6}$IPAC, California Institute of Technology, Mail Code 314-6, 1200 E. California Blvd., Pasadena, CA 91125, USA 
}

\date{Accepted XXX. Received YYY; in original form ZZZ}

\pubyear{2020}

\begin{document}
\label{firstpage}
\pagerange{\pageref{firstpage}--\pageref{lastpage}}
\maketitle

\begin{abstract}

The spatial distribution of massive and luminous galaxies have provided important constraints on the fundamental cosmological parameters and physical processes governing galaxy formation. In this work, we construct and compare independent galaxy-halo connection models in the application of clustering measurement at non-linear scales of BOSS-CMASS galaxies. In particular, we adopt a halo occupation distribution (HOD) model with 11 parameters and a semi-analytical model (SAM) with 16 parameters to describe the galaxy two point correlation function. With an empirical parameterization for the velocity field to model the redshift space distortion effect and the emulator technique, we can explore the parameter space of both models. We find that the HOD model is able to recover the underlying velocity field of SAM with an accuracy of 3\%, and can be improved to 1\% when the analysis is restricted to scales above 1$h^{-1}$Mpc. The comparison is based on multiple samplings in the parameter space which can verify the convergence of the models. Then we perform constraints on the model parameters using clustering measurement of CMASS galaxies. Although limited by the emulator accuracy and the flexibility of the model, we find that the clustering measurement is capable of constraining a subset of the SAM parameters, especially for components sensitive to the star formation rate. This result leads us to anticipate that a joint analysis of both clustering and abundance measurements can significantly constrain the parameters of galaxy formation physics, which requires further investigation from both theoretical and observational aspects. 

\end{abstract}


\section{Introduction}

Galaxy formation and evolution are highly non-linear and non-trivial processes that couples numerous astrophysical, radiative, cosmological, and dynamical processes. Due to developments from both theoretical and observational perspectives, our understanding of galaxy formation has achieved significant advances in the past few decades. However, there are still substantial unknowns and gaps in understanding for a complete theory of this topic (\citealt{Somerville_2015, Naab_2017, Wechsler_2018, Forster_Schreiber_2020, Lapi_2025}). One of the major methods is to investigate the statistical properties of the galaxy population and infer the underlying parameters with a given set of theoretical and physical assumptions (\citealt{Romeo_2020, Romeo_2023}). The spatial distribution of galaxies is believed to contain a significant amount of information, as demonstrated by large-scale galaxy surveys such as the Sloan Digital Sky Survey (SDSS-I/II, \citealt{SDSS_York, Abazajian_2009}), the Two-degree Field Galaxy Redshift Survey (2dFGRS, \citealt{Colless_2001, Cole_2005}), WiggleZ (\citealt{Drinkwater_2010}), SDSS-BOSS (\citealt{Dawson_BOSS}), SDSS-eBOSS (\citealt{eBOSS_Dawson}), the Dark Energy Spectroscopic Instrument (DESI, \citealt{DESI_2016, DESI_2024_BAO}) and so on. The spatial distribution, characterized by the clustering measurements, is sensitive to various physical processes, especially at non-linear scales, where both the cosmological model and the galaxy formation physics matter. In order to fully extract the information from the clustering measurement, it is necessary to build a unbiased connection between galaxies and their host dark matter halos, see e.g. \cite{Wechsler_2018} for a detailed review and references therein. 

The halo occupation distribution (HOD) is an empirical model widely used to interpret the clustering measurements of different types of galaxies due to its simplicity and flexibility (\citealt{Jing_1998, Benson_2000, HOD_Weinberg, Zheng_2005}). In this framework, the key input is the probability distribution that a dark matter halo of mass $M$ contains $N$ galaxies. Combined with some prescription of the phase space distribution of these galaxies within their halos and secondary parameters beyond $M$, this model is quite successful in the clustering analysis and can provide basic profile information such as the satellite fraction of the galaxy sample, the average mass of the host halos, linear bias, and so on. Therefore, we can combine this empirical model with an accurate N-body simulation to fully explore the correlation between galaxy distribution and cosmological models. In \cite{Zhai_2019, Zhai_2023a}, we proceeded along this direction with an emulator approach to quickly and accurately predict the clustering signals for an arbitrary parameter set of both cosmology and HOD. The output reveals substantial constraining power on the fundamental cosmological parameters using information at highly non-linear scales, as also demonstrated by others (\citealt{Wibking_2017, Yuan_2022, Chapman_2021, Salcedo_2022, Lange_2023}). 

On the other hand, physical models such as hydrodynamical simulation or semi-analytical models (SAM) choose a different route by directly simulating or modeling the basic processes in galaxy formation (\citealt{Benson_2010}). Compared with hydrodynamical simulation, SAM requires much fewer computational resources to reach sufficient volume and resolution. However, SAM incorporates many necessary components with a large number of free parameters, and a sufficient exploration in the parameter space can be challenging. Thus the extraction of the underlying cosmological parameters becomes even more difficult than in the case of HOD.

In this work, we aim to perform a clustering analysis using SAM for galaxy formation. The process is similar to the earlier works for massive galaxies (\citealt{Parejko_LOWZ, CMASS_Martin, Zhai_2017}) but with the key component HOD replaced by SAM. We note that there are multiple motivations for us to do such an analysis. First, in order to extract the cosmological information using galaxy clustering at non-linear scales with minimal bias, the uncertainties in the modeling of galaxy-halo connection should be carefully examined and statistically marginalized. The HOD and SAM approaches are separate methods with distinct assumptions about the underlying physics --- using different methods in the cosmological data analysis enables a cross-check of the results and provides new insights. In terms of the cosmological measurement expressed as the linear growth rate $f\sigma_{8}$, the HOD-based method leads to a somewhat lower value compared to the prediction of the Planck experiment (\citealt{Zhai_2023a}). Therefore, an analysis with an independent method, i.e., SAM, can shed new light on the result. From the perspective of galaxy formation, HOD and SAM can describe the galaxy sample in different ways, but there are parameters and properties that can be predicted by both, for instance, the satellite fraction. This can provide a reference point to evaluate the consistency and inconsistency of the models. 

Second, in addition to the comparison of the empirical and physical models, we note that a SAM typically has more parameters in the model. This is not surprising since it takes more components (those from galaxy formation physics) into account instead of only focusing on the clustering property. Therefore, the search for a reasonable set of SAM parameters is of critical importance and is also challenging. This calibration process is not a new problem in the study of SAMs; there has been progress made in many previous works by exploring the parameter space with an emulator of galaxy statistics (\citealt{Bower_2010, Ruiz_2015, Oleskiewicz_2020, Elliott_2021}), Bayesian model inference in high dimensional space (\citealt{Lu_2011, Lu_2012, Lu_2014}), examination of model prediction with observational data at low and high redshifts (\citealt{Benson_2014, Knebe_2018, Henriques_2020}) and so on. The output of this process is the calibrated set of parameters in a quantitative manner. We note that in most of the studies, the calibration or the search of parameter sets utilizes only abundance information, or one-point statistics, such as the luminosity function in certain wavelength band, or stellar mass function in some redshift range. There are far fewer works focusing on the calibration using two-point statistics. This is not surprising since the clustering typically probes scales beyond those that are sensitive to the modeling in SAMs. However, state-of-the-art observational data have been able to provide accurate clustering measurement down to scales below $1h^{-1}$Mpc. Although this may still be larger than many baryonic processes governing galaxy formation, there could be some influence on the spatial distribution of massive galaxies, as we explore in this paper. Indeed, a few recent works show that the addition of the two-point statistics can provide another enhancement of model constraints, including both galaxy clustering and galaxy lensing (\citealt{vanDaalen_2016, Renneby_2020}). Therefore, it motivates us to include a clustering analysis in parallel with the HOD approach, and we expect to obtain constraints on some of the SAM modeling parameters.

Compared with the widespread applications of the HOD-based method in the study of galaxy clustering, the potential power of SAMs has not been fully exploited. Part of the reason is that the HOD model can be easily adopted on a N-body simulation with a sufficiently large volume. However, the computing demands of SAMs to process all the merger trees of dark matter halo in the same simulation is orders of magnitude higher and thus makes the computation not feasible. One possible solution is to adopt an analytical method based on Monte Carlo and Extended Press-Schecter theory (\citealt{Bond_1991, Bower_1991, Lacey_1993}) or calibrated with numerical simulations (\citealt{Parkinson_2008}), and this method has been widely used in previous works. Although fast, this method is almost purely determined by halo mass and does not have accurate information about the phase space distribution of dark matter halos and thus prevents the application of SAM to study galaxy clustering in an accurate manner. To solve the problem, we use the merger trees of dark matter halos from the simulation in this analysis. Since we are modeling massive galaxies as in BOSS-CMASS, this allows us to focus on the massive branches and make the necessary approximations to simplify the computation. To enable a fair comparison, we use the same simulation for both HOD and SAM-based analysis in this work.

In order to fully compare the HOD and SAM models in terms of both cosmology and galaxy formation, the ideal framework needs a simulation suite similar to Aemulus (\citealt{DeRose_2018}) that can cover a sufficiently large volume in the cosmological parameter space. However, such simulations are not available for a variety of reasons, for instance, the lack of merger trees of dark matter halos. Therefore, we move one step back by starting from one single simulation with both dark matter halos and merger trees in this work. We will leave the analysis and generation of a simulation suite with varying cosmological model and dark matter halo merger trees to a future work.

Our paper is organized as follows: Section \ref{sec:simulation} describes the simulation and data we use. Section \ref{sec:method} introduces both the HOD and SAM models we compare throughout the analysis. Section \ref{sec:results} presents the results. We discuss and conclude in Section \ref{sec:conclusion}.

\section{Simulation and data}\label{sec:simulation}

To build the HOD and SAM models for galaxy clustering analysis, we adopt the UNIT\footnote{\url{https://unitsims.ft.uam.es}} \citep{Chuang_2019} simulation throughout this work. This simulation is run with a Planck 2015 cosmology (\citealt{Planck_2015}), the box size is $1h^{-1}$Gpc and the total number of dark matter particles is $4096^{3}$. This volume and particle mass are able to resolve dark matter halos for CMASS-like galaxies. The simulation adopts the inverse phase technique (\citealt{Angulo_2016}) to reduce cosmic variance and has two pairs (4 boxes) as the final output. Using all these four boxes will have better variance for the summary statistics of both dark matter halos and galaxies. However, the computation of the SAM is still computationally expensive, thus we limit ourselves to using only one of the boxes fixedAmp\_001, and we anticipate that this does not significantly impact our final results for the comparison of different galaxy-halo connection models. 

The dark matter halos are identified using the publicly available ROCKSTAR halo finder (\citealt{Behroozi_2013b}). The merger trees as input for SAM are constructed using Consistent Tree software (\citealt{Behroozi_2013}). In order to have accurate modeling of galaxy statistics using SAM, we do not apply any downsampling of the simulation to use only a subvolume or subset, but instead, we use all the 160 million merger trees from the UNIT simulation box. 

The observational dataset we plan to use is from the large scale structure catalog from BOSS observations (\citealt{Reid_2016}). Following the earlier works in \cite{Zhai_2023b, Zhai_2023a}, we focus on the galaxy sample at $z=0.55$ selected from the CMASS compilation and apply the magnitude cut on the sample to have a more complete sample for clustering measurement. The galaxy mocks based on the UNIT simulation are constructed using the $z=0.556$ snapshot to best match the CMASS redshift.

\section{Methodology}\label{sec:method}

In this section, we introduce the analysis pipeline, including the observables for galaxy clustering, the galaxy-halo connection model for both HOD and SAM, and the statistical inference method for the model constraints.

\subsection{Clustering observables}

We use the two-point correlation function (2PCF) $\xi(r)$ to quantify the clustering signals for both the observational datasets and simulations. By definition, it measures the excess probability of finding two galaxies separated by a vector distance $\mathbf{r}$ relative to a random distribution for all $|\mathbf r| = r$. In the calculation, we can decompose the galaxy pairs and measure $\xi(r)$ on a two-dimensional grid. One of the decompositions is to project the separation along the line-of-sight ($\pi$) and perpendicular ($r_{p}$) to the line-of-sight to have $\xi(\pi, r_{p})$. In order to mitigate the redshift space distortion effect due to peculiar velocity in the $\pi$ direction and obtain clustering amplitude in real space, we compute the projected correlation function
\begin{equation}
w_{p}(r_{p})=2\int_{0}^{\infty}d\pi\,\xi(r_{p}, \pi).
\end{equation}
In practice, we need to truncate the integral at some scale using data from observation or simulation. We choose $\pi_{\mathrm{max}}=80h^{-1}$Mpc throughout our analysis to be consistent with earlier works.

Another way to decompose the separation in the measurement is via $s^{2}=\pi^{2}+r_{p}^2$ and $\mu=r_{p}/s$ to have $\xi(s, \mu)$. It can incorporate the clustering in redshift space and enable the measurement of Legendre multipoles through
\begin{equation}
\xi_{\ell}(s) =\frac{2\ell+1}{2}\int_{-1}^{1} L_{\ell}(\mu)\,\xi_{Z}(s, \mu)d\mu,
\end{equation}
where $L_{\ell}$ is the Legendre polynomial of order $\ell$. Adding higher orders of multipoles into the analysis can have additional constraining power but increasing contamination due to noise, therfore we only focus on the first two orders: $\xi_{0,2}$. 

We measure the 2PCF from the BOSS galaxy sample through the estimator (\citealt{LS_1993})
\begin{equation}\label{eq:LS}
\xi(r_{p}, \pi)=\frac{DD-2DR+RR}{RR},
\end{equation}
where DD, DR and RR are the number of pairs of data-data, data-random and random-random in each scale bins. For the galaxy mocks generated from simulation, we use the natural estimator for simplicity
\begin{equation}
    \xi_{r_{p}, \pi} = 1-\frac{DD}{RR}.
\end{equation}
Throughout the analysis for both BOSS galaxies and mocks, we choose nine logarithmic bins for both $r_{p}$ and $s$ in the range from 0.1 to 60.2 $h^{-1}$Mpc. For 2PCF multipoles, $\mu$ is binned linearly with 40 bins from 0 to 1.

\subsection{Galaxy-halo connection models}

\begin{table}
\centering
%\begin{tabularx}{0.5\textwidth}{|p{1cm}|>{\raggedright\arraybackslash}X|>{\raggedright\arraybackslash}X|}
\begin{tabularx}{0.5\textwidth}{m{1cm}|>{\centering\arraybackslash}X|m{1.5cm}}
\hline
Parameter    & Meaning & Range \\
\hline
$\log{M_{\rm{sat}}}$  & The typical mass scale for halos to host one satellite on average & [14.0, 15.5] \\\hline
$\alpha$  & The power-law index for the mass dependence of the number of satellites & [0.2, 2.0]\\\hline
$ \log{M_{\rm{cut}}}$ & The mass cut-off scale for the satellite occupation function  &  [10.0, 13.7]\\\hline
$\sigma_{\log{M}}$  & The scatter of halo mass at fixed galaxy luminosity  & [0.05, 1.0]\\\hline
$\eta_{\rm{con}}$ & The concentration of satellites relative the dark matter halo   &  [0.2, 2.0]\\\hline
$\eta_{\rm{vc}}$  &  The velocity bias for central galaxies  &   [0.0, 0.7]\\\hline
$\eta_{\rm{vs}}$  & The velocity bias for satellite galaxies  &   [0.2, 2.0]\\\hline
$f_{\text{max}}$ & The incompleteness parameter for central occupancy & [0.1, 1.0] \\ \hline
$f_{\rm{env}}$  & Amplitude parameter for assembly bias  &   [-0.3, 0.3]\\\hline
$\delta_{\rm{env}}$  & Position parameter for assembly bias  &   [0.5, 2.0]\\\hline
$\sigma_{\rm{env}}$  & Width parameter for assembly bias  &   [0.1, 1.0]\\
\hline\hline
$\gamma_{f}$  & Amplitude of halo velocity field relative to the raw simulation output  &   [0.5, 1.5]\\\hline
$\bar{n}$  & number density of galaxy sample in unit of $10^{-4}$/$[h^{-1}$Mpc]  &   [1.5, 4.5]\\
\hline
\end{tabularx}
\caption{The first section shows the parameters of our HOD model, including their physical meaning and the range of values explored. The prior in the likelihood analysis is the same as the range listed here, i.e. uniform and flat without extrapolation. The second section shows two additional parameters that are also used for SAM modeling. See text for more details.}
\label{tab:HODparam}
\end{table}

\begin{table}
\centering
%\begin{tabularx}{0.5\textwidth}{|p{1cm}|>{\raggedright\arraybackslash}X|>{\raggedright\arraybackslash}X|}
\begin{tabularx}{0.5\textwidth}{m{1cm}|>{\centering\arraybackslash}X|m{1.5cm}}
\hline
Parameter    & Meaning & Range \\
\hline
$\nu_{\rm{SF,0}}$  & normalized frequency factor for star formation rate in disk  & [1e-10, 3e-8]\\\hline
$\epsilon_{\star}$  & efficiency parameter for star formation in the spheroid & [0.01, 1]\\\hline
$ \alpha_{\star}$ & exponent parameter for star formation in the spheroid &   [-3, 3]\\\hline
$\beta_{\mathrm{max, d}}$  & maximum of the mass loading factor in the disk  &  [100, 1000]\\\hline
$V_{\mathrm{outflow, d}}$ & velocity of the outflow in the disk   &  [10, 300] \\\hline
$\alpha_{\mathrm{outflow, d}}$  &  index of the outflow model in the disk  &   [1, 4]\\\hline
$\beta_{\mathrm{max, s}}$  & maximum of the mass loading factor in the spheroid  & [100, 1000]\\\hline
$V_{\mathrm{outflow, s}}$ & velocity of the outflow in the spheroid   &  [10, 300]\\\hline
$\alpha_{\mathrm{outflow, s}}$  &  index of the outflow model in the spheroid  &  [1, 4]\\\hline
$f_{\mathrm{core}}$  & core radius of the hot halo in units of the virial radius of the host dark matter halo &  [0.01, 0.8] \\\hline
$f$ & age factor for the computation of time available for cooling &  [0, 1] \\ \hline
$c_{\mathrm{multi}}$  & multiplier factor to scale the cooling rate from the raw computation &  [0.1, 10.0] \\\hline
$\gamma$  & $\gamma$ in the parameterized model for outflow reincorporation (Eq. \ref{eq:rein})  &  [0.05, 20.0]\\\hline
$\delta_{1}$  & $\delta_{1}$ in the parameterized model for outflow reincorporation (Eq. \ref{eq:rein})  &   [0.3, 5.0]\\\hline
$\delta_{2}$  & $\delta_{2}$ in the parameterized model for outflow reincorporation (Eq. \ref{eq:rein})  &  [0.3, 5.0]\\\hline
$\eta_{\mathrm{radio}}$  & efficiency parameter for the AGN feedback due to radio-mode  &  [0.01, 3.0]\\\hline
\hline\hline
$\gamma_{f}$  & Amplitude of halo velocity field relative to the raw simulation output  &   [0.5, 1.5]\\\hline
$\bar{n}$  & number density of galaxy sample in unit of $10^{-4}$/$[h^{-1}$Mpc]$^3$  &   [1.5, 4.5]\\
\hline
\end{tabularx}
\caption{The same as Table \ref{tab:HODparam} but for the Galacticus SAM.}
\label{tab:SAMparam}
\end{table}

The key ingredient of our clustering analysis at highly non-linear scales is the galaxy-halo connection model. We introduce these models in more detail in this section, including the HOD model as an empirical model and SAM as a physical model. Their main features and parameters are summarized in Table \ref{tab:HODparam} and \ref{tab:SAMparam}.

\subsubsection{Halo occupation distribution}

The key component in the HOD modeling is the probability distribution $P(N/M)$ that a dark matter halo of mass $M$ contains $N$ galaxies of some type. In the most basic form, this depends only on $M$ but the flexibility of this model allows the implementation of more components in a straightforward manner. In this parameterization, we separate the modeling into central and satellite galaxies as 
\begin{equation}\label{eq:NM}
    \langle N (M)\rangle = \langle N_{\mathrm{cen}} (M)\rangle + \langle N_{\mathrm{sat}} (M)\rangle.
\end{equation}
The mean occupation of central galaxies is 
\begin{equation}
  N_{\text{cen}}(M) = \frac{1}{2} \left[1+\text{erf} \left(\frac{\log_{10}{M}-\log_{10}{M_{\text{min}}}}{\sigma_{\log{M}}}\right)\right],  
\end{equation}\label{N_cen}
while for satellite galaxies, it is
\begin{equation}
  N_{\text{sat}}(M) = \left(\frac{M}{M_{\text{sat}}}\right)^{\alpha}\exp{\left(-\frac{M_{\text{cut}}}{M}\right)} N_{\text{cen}}(M).  
\end{equation}
In the modeling of individual halos, the number of central and satellite galaxies are random draws from Bernoulli and Poisson distributions, respectively. This model follows the study first proposed in \cite{Zheng_2005}, which has been widely used in the literature, and we add a few more parameters for additional calibration. We introduce the components to fully define our model as follows

\textbullet~ \textbf{The basic parameters}: The parameters of this model include $M_{\mathrm{min}}, \sigma_{\log{M}}, \alpha, M_{\mathrm{cut}}$ and $M_{\mathrm{sat}}$. Typically, $M_{\mathrm{min}}$ approximates the mass at which the average occupation of the central galaxy is 0.5, 
$\sigma_{\log{M}}$ relates to the scatter of the halo mass at fixed galaxy luminosity, $\alpha$ is the power-law index for the mass dependence of satellite occupation at the massive end, $M_{\mathrm{cut}}$ is an additional cutoff for satellite occupation, and $M_{\mathrm{sat}}$ is a typical mass for halos to host one satellite. This set of parameters can fully specify the HOD model, and their value can be obtained by fitting with observations. When we apply this HOD model to the CMASS galaxies, the number density of the sample can serve as an additional constraint via
\begin{equation}
    \bar{n} = \int \frac{\mathrm{d}n}{\mathrm{d}M}N(M),
\end{equation}
where $\mathrm{d}n/\mathrm{d}M$ is the halo mass function and $N(M)$ is the HOD from Eq.(\ref{eq:NM}). We use the measurement of $\bar{n}$ from the sample and calculate $M_{\mathrm{min}}$ as a derived parameter instead of a free parameter when the other HOD parameters are known.

\textbullet~ \textbf{The additional phase space parameters: $\eta_{\mathrm{con}}, \eta_{\mathrm{vs}}$ and $\eta_{\mathrm{vc}}$}. The dark matter distribution within the halo can be well described by the NFW profile (\citealt{NFW_1996}), and the satellite distribution can differ from it. We introduce the parameter $\eta_{\mathrm{con}}=c_{\mathrm{sat}}/c_{\mathrm{halo}}$ to scale the concentration of the satellite distribution compared with dark matter. $\eta_{\mathrm{vs}}$ and $\eta_{\mathrm{vc}}$ are the velocity bias parameters for satellite and central galaxies, respectively (\citealt{Guo_2015}). They can scale the velocity of galaxies relative to the host halo based on the velocity dispersion $\sigma_{\mathrm{halo}}$.

\textbullet~ \textbf{Assembly bias parameters: $f_{\mathrm{env}}, \sigma_{\mathrm{env}}$ and $\delta_{\mathrm{env}}$:} Numerical simulation shows that the clustering properties of dark matter halos and the galaxy occupation can be affected by parameters other than the halo mass (\citealt{Gao_2005, Wechsler_2006}). We can model this effect within the HOD approach based on the external environment (\citealt{Han_2019, Xu_2020, Yuan_2021b}). In order to do so, we measure the relative density of the dark matter halo using a top-hat window function of radius $\sim10h^{-1}$Mpc (\citealt{Mcewen_2018}). Then we modify the HOD parameter $M_{\mathrm{min}}$ as
\begin{equation}\label{eq:Mmin_AB}
    \bar M_{\rm min} = M_{\text{min}}\left[1 + f_{\rm env}\,\text{erf}\left(\frac{\delta-\delta_{\rm env}}{\sigma_{\rm env}}\right)\right].
\end{equation}
In addition to the constraint from the galaxy number density, this parameterization can modulate the dependence of $M_{\mathrm{min}}$ based on the local density. The parameter $f_{\mathrm{env}}$ controls the overall sensitivity of galaxy clustering to the effect of assembly bias, $\delta_{\mathrm{env}}$ and $\sigma_{\mathrm{env}}$ work as additional degrees of freedom to model the transition of halos from the under-dense to over-dense regions. We note that the choice of this external environment as the secondary parameter for clustering analysis is not unique due to the lack of a tight constraint on this assembly bias effect from observations. The HOD model allows exploration of this effect using both internal and external properties with substantial flexibility, see e.g. \cite{Hearin_2016, zu_mandelbaum:16}.

\textbullet~\textbf{Incompleteness parameter: $f_{\mathrm{max}}$.} The basic HOD form assumes that the central occupation approaches unity at the massive end. However, the target selection and mass incompleteness (\citealt{leauthaud_etal:16}) of the BOSS galaxies show that this is not necessarily true, i.e. some of the most massive dark matter halos may not have a BOSS galaxy in their center. To consider this effect, we multiply $N_{\mathrm{cen}}$ by $f_{\mathrm{max}}$ in the HOD modeling and allow $f_{\mathrm{max}}$ to be fitted by data.

\subsubsection{Semi-analytical model: Galacticus}

There are many semi-analytical models for galaxy formation developed and calibrated for different purposes in the community, but most of them work in a similar fashion. In this work, we use the Galacticus SAM (\citealt{Benson_2012}) as a physical model to describe the galaxy-halo connection. Due to the high dimensional parameter space, it is not feasible to fully explore all the components in SAM. We thus focus on a subset of them and introduce the salient features. More technical details can be found in the summary in a recent work by \cite{Wang_2022}.

\textbullet~\textbf{Star formation:} We model the star formation in the disc using the prescription in \cite{Blitz_2006}. The star formation rate is given by $\dot{\Sigma}_{\star}(R)=\nu_{\mathrm{SF}}(R)\Sigma_{\mathrm{H}_{2},\mathrm{disk}}(R)$, where $\Sigma_{\mathrm{H}_{2},\mathrm{disk}}(R)$ is the surface density of molecular gas and $\nu_{\mathrm{SF}}$ is the frequency depending on the surface density of hydrogen in both molecular and atomic forms. We choose the normalization factor of $\nu_{\mathrm{SF,0}}$ as a free parameter in our SAM modeling. The star formation rate in the galactic spheroid is calculated by dividing the mass of gas in the spheroid by a star formation timescale. Specifically, the time scale is computed as $\tau_{\star}=\epsilon_{\star}^{-1}\tau_{\mathrm{dynamical}}(V/200\,\mathrm{km\,s^{-1}})^{\alpha_{\star}}$, where $\tau_{\mathrm{dynamical}}=r/V$ is the dynamical timescale of the spheroid, $r$ and $V$ are the characteristic radius and velocity. The parameter efficiency $\epsilon_{\star}$ and exponent $\alpha_{\star}$ are allowed to vary in our model.

\textbullet~\textbf{Stellar feedback:} We model the stellar feedback in the disc and spheroid independently but using the same parameterization. We allow the maximum of the mass loading factor $\beta_{\mathrm{max}}$ to vary. In addition, the outflow rate is a simple power law model given by 
\begin{equation}
    \dot{M}_{\mathrm{outflow}}=\left(\frac{V_{\mathrm{outflow}}}{V}\right)^{\alpha_{\mathrm{outflow}}}\frac{\dot{E}}{E_{\mathrm{canonical}}},
\end{equation}
where $V$ is the characteristic velocity of the component (i.e., disk or spheroid), $\dot{E}$ is the rate of energy input from stellar populations, and $E_{\mathrm{canonical}}$ is the total energy input by a normalized canonical stellar population. The other parameters, $V_{\mathrm{outflow}}$, the velocity of the outflow, and $\alpha_{\mathrm{outflow}}$, the index of the dependence, are input parameters allowed to vary.

\textbullet~\textbf{Cooling:} We use the prescription from \cite{White_1991} to calculate the cooling rate from the hot halo. In this model, one of the parameters is denoted as $f_{\mathrm{core}}$, the core radius of the hot halo in units of the virial radius of the dark matter halo. The hot halo mass distribution is a spherically symmetric $\beta-$profile
\begin{equation}
    \rho_{\mathrm{hothalo}}(r)\propto[r^{2}+r_{\mathrm{core}}^{2}]^{3\beta/2},
\end{equation}
where the core radius $r_{\mathrm{core}}$ is determined via $f_{\mathrm{core}}$. The time available for cooling is equal to
\begin{equation}
    t_{\mathrm{available}} = \mathrm{exp}[f\ln{t_{\mathrm{Universe}}}+(1-f)\ln{t_{\mathrm{dynamical}}}],
\end{equation}
where $t_{\mathrm{Universe}}$ is the age of the Universe, $t_{\mathrm{dynamical}}$ is the dynamical time of the halo, and $f$ is a modeling parameter that can interpolate between $t_{\mathrm{Universe}}$ and $t_{\mathrm{dynamical}}$. To enable further flexibility, we introduce another parameter $c_{\mathrm{multi}}$ to multiply the cooling rate output from the above model as the final result for mock galaxies.

\textbullet~\textbf{Outflow reincorporation:} Gas outflowing from the galaxy is retained in a reservoir, but can be gradually reincorporated into the hot halo. We adopt the model from \cite{Henriques_2013} for this process. This model assumes that the gas returns at a rate
\begin{equation}\label{eq:rein}
    \dot{M}_{\mathrm{return}} = \gamma(1+z)^{-\delta_{1}}\left(\frac{V_{\mathrm{vir}}}{200\,\mathrm{km/s}}\right)^{\delta_{2}}\frac{M_{\mathrm{outflowed}}}{{\tau_{\mathrm{dynamical}}}},
\end{equation}
where $V_{\mathrm{vir}}$ is the virial velocity of the halo, $M_{\mathrm{outflowed}}$ is the current amount of gas in the reservoir, the parameters $\gamma, \delta_{1}$ and $\delta_{2}$ are allowed to vary in our model.

\textbullet~\textbf{AGN feedback:} The supermassive black hole in the center of the galaxy can accrete from both the hot gas halo and the interstellar medium of the spheroid. We implement both quasar-mode and radio-mode feedback in the model. The quasar-mode feedback allows the black hole to drive a wind from the spheroid, while the radio-mode feedback encapsulates the jet power deposition into the hot halo. For simplicity, we only allow the parameter $\eta_{\mathrm{radio}}$ for radio-mode to vary, which defines the efficiency that the feedback is coupled to the circum-galactic medium (CGM).

\subsubsection{Additional parameters}

Our analysis is based on a single cosmological simulation UNIT, without fully sampling the cosmological parameter space. We note that some approximate methods, such as the rescaled N-body simulation combined with a SAM, have been used to constrain parameters like $\sigma_{8}$ (\citealt{Harker_2007}). In this work, in order to extract the cosmological information from galaxy clustering at non-linear scales, we rely on the RSD effect due to peculiar velocities. We apply a parameter $\gamma_{f}$ as first introduced by \cite{Reid_2016}, to scale the velocity field of dark matter halos from the simulation. It is based on the approximation that a fractional change of $\gamma_{f}$ can measure the linear growth rate $f\sigma_{8}$. This parameter has already been adopted in earlier works (\citealt{Zhai_2023a, Chapman_2021}) to yield robust measurement and extended in \cite{Chapman_2023} to split the effect and modeling at both linear and non-linear scales. In this work, the introduction of this parameter can significantly increase the range of $f\sigma_{8}$ anchored at the simulation cosmology and equivalently enable constraint using observational data. We adopt this parameter in both the HOD model and Galacticus SAM.

Although our simulation focuses on the CMASS galaxies and the number density of the sample provides an additional constraint, we can make the model more generalized. We do not produce galaxy mocks at fixed number density, but make $\bar{n}$, the galaxy number density a parameter in the range [1.5e-4, 4.5e-4] in units of [$h^{-1}$Mpc]$^{-3}$. This can cover the range of CMASS sample, and subsamples with various selections.

\subsection{Analysis methodology}

The observable in our analysis is the 2PCF of galaxies. It is straightforward to compute in the HOD model since its output is mainly the phase space information of the galaxy mock. On the other hand, the output from SAM has many more parameters in addition to position and velocity. To mimic the selection of CMASS galaxies, we can apply the same cut on magnitude and color on the galaxy mock from SAM since it can incorporate the star formation history and stellar population model to predict the spectrum and luminosity at a given wavelength band. However, it is nontrivial for the resultant galaxy mock to match the observations for all the properties, such as color cuts and number density. In order to simplify the analysis, we use the stellar mass predicted from Galacticus for our selection with rank ordering. We could adopt a similar strategy to select galaxies based on luminosity and color. However, this is not straightforward to achieve, since the cuts on multiple parameters can lead to very different number densities for different SAM models. On the other hand, previous results show that the CMASS galaxies are likely to be red, massive galaxies in massive dark matter halos (\citealt{Anderson_2012}). Measurements of their luminosity function and stellar mass function indicate that the sample is more complete at the massive end of stellar mass (\citealt{leauthaud_etal:16, Bernardi_2016}). Furthermore, HOD analysis using clustering measurements show that these galaxies preferentially live in more massive halos (\citealt{CMASS_Martin}) and it is more likely for them to be more massive in stellar mass as well and match the selection of SAM mocks. In particular, we rank-order the SAM galaxies by stellar mass and select from the most massive until we reach the required number density. We note that with improved predictions of color and magnitude, it would be worthwhile to compare the selection of SAM mocks and observational data in a more accurate manner.

In order to explore the parameter space of both HOD and SAM by fitting the 2PCF, we need to make the theoretical prediction of a given parameter set with sufficient accuracy and speed. In order to solve this problem, we follow the previous work in \cite{Zhai_2023a} to build an emulator based on Gaussian Processes (\citealt{RW_2006}). It has shown that their application to HOD-based models is quite effective, even when the cosmological parameters are also allowed to vary. A few thousand training samples are able to make the prediction accurate to a few percent on the scales of interest. In this work, the application to the single cosmology is simpler and straightforward. Roughly 3,000 samples or fewer can fulfill our requirement of both accuracy and speed. On the other hand, it is non-trivial for SAMs since the computational cost of Galacticus is orders of magnitude higher than a HOD. It is not feasible to run thousands of SAM on the UNIT simulation and, for each model, fully process all the 160 million merger trees. Given the concern of speed and the fact that the CMASS galaxies are likely to live in very massive dark matter halos, we do not process branches of the merger trees corresponding to halos of mass less than $10^{11}\mathrm{M}_\odot$. Furthermore, we prune all branches from the tree corresponding to halo masses below $10^{10}\mathrm{M}_\odot$. This is not ideal since it is likely that a tree with branches below the limit can also evolve to have a CMASS-like galaxy, but it can improve the efficiency significantly, although still not comparable to the HOD-based model. Considering the balance between accuracy and speed, we finally run roughly 500 Galacticus models, and each processes the whole UNIT simulation box \footnote{The sampling in the parameter space of both the HOD and SAM model is based on a Latin hypercube design. The initial design for Galacticus SAM uses 1,000 points, but it turns out that some models with certain parameter combinations of $\nu_{\mathrm{SF,0}}$ and $c_{\mathrm{multi}}$ can yield unphysically high star formation and make the solution of differential equation extremely expensive, therefore we exclude these models in our final production and proceed with the 500 completed models.}. The resultant accuracy of the 2PCF emulator is not comparable to the HOD model due to not only the fewer training samples but also the higher dimensional parameter space, which makes the sampling sparser. 

\begin{figure*}
\begin{center}
\includegraphics[width=17cm]{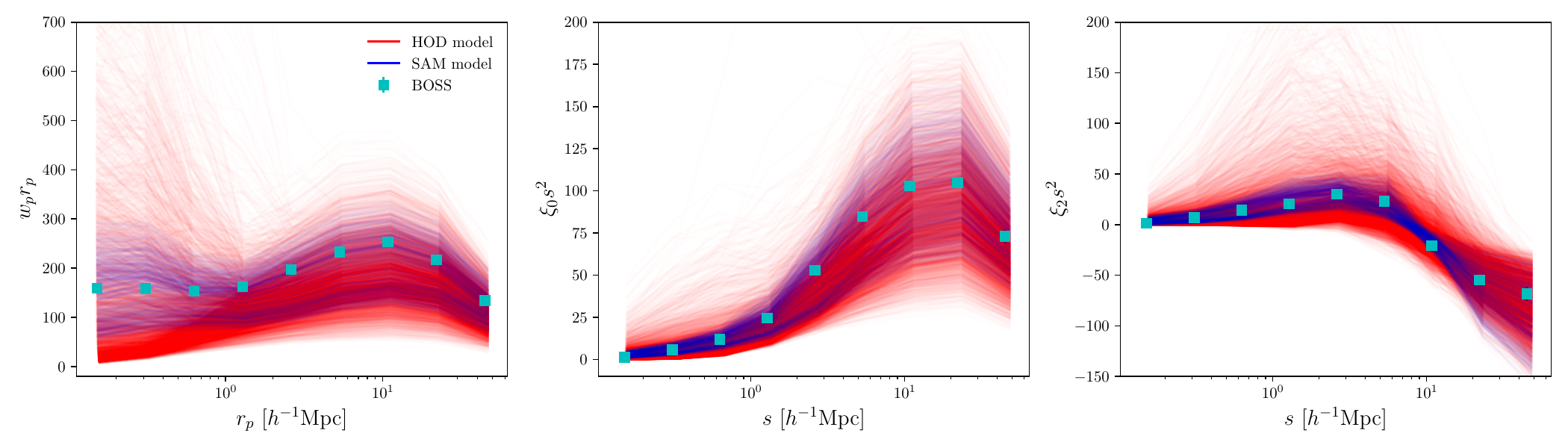}
\caption{The 2-point correlation function ($w_{p}, \xi_{0}$ and $\xi_{2}$) predicted from the HOD and SAM model, respectively, compared with the measurement from BOSS-CMASS galaxy sample. The HOD model has 4000 sampling in the parameter space, while the Galacticus SAM has 500 sampling. It is clear that the range of the parameters of both models can produce clustering results to encompass the observed measurement.}
\label{fig:all_2pcf}
\end{center}
\end{figure*}

\begin{figure*}
\begin{center}
\includegraphics[width=17cm]{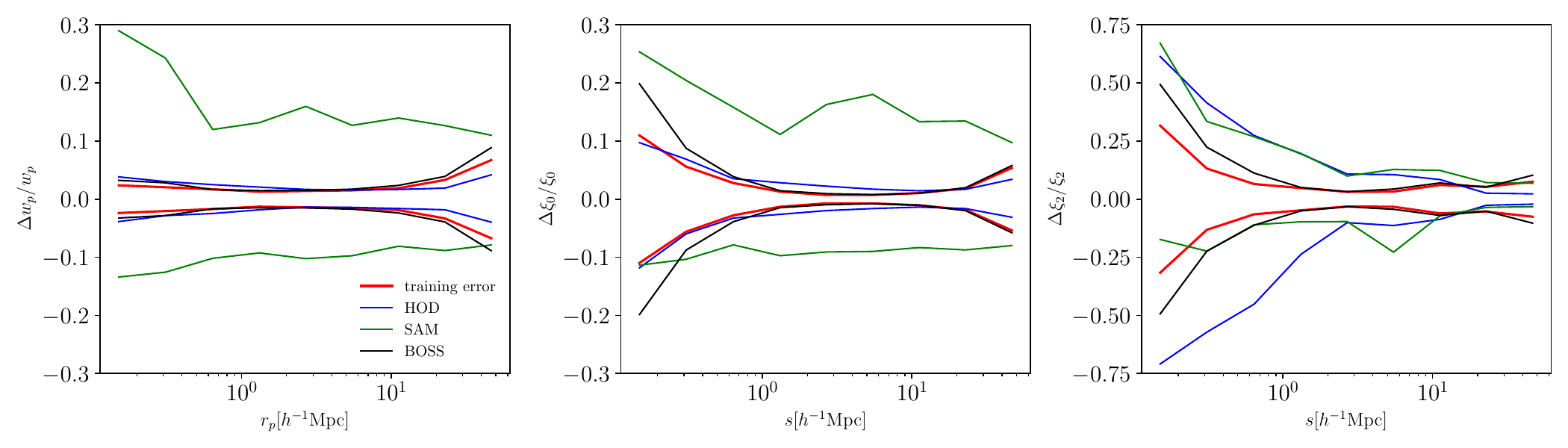}
\caption{The GP-based emulator error of the two-point correlation function using HOD and SAM models, respectively. The HOD model uses 80\% of the mocks for training, while SAM model uses 90\% for training, and the rest are used to quantify the emulator error. The lines show the inner 68\% of the fractional difference from all the test samples. For comparison, the input training error and the uncertainty from BOSS galaxies are also shown.}
\label{fig:emulator_error}
\end{center}
\end{figure*}

In Figure \ref{fig:all_2pcf}, we show the 2PCF predicted from the HOD and SAM mocks. The HOD model has 4,000 samplings in the parameter space, while the Galacticus SAM has 500 samplings. The result shows that these mocks can produce clustering measurements to enclose the CMASS measurements and enable data fitting without extrapolating the parameter space. Based on these mocks, we use 80\% of the HOD mocks and 90\% of the SAM mocks as the training data to build the emulator of 2PCF, and use the rest to test the accuracy of the emulator. As Figure \ref{fig:emulator_error} shows, the HOD-based model is quite accurate as in previous works. For $w_{p}$ and $\xi_{0}$, the overall accuracy is at the percent level, and is comparable to or lower than the sample variance. On the other hand, the SAM based model only reaches a level of roughly ten percent, and is more dominant compared with the uncertainty from observations. The result for $\xi_{2}$ is slightly different, the accuracy of the SAM based model is comparable to or better than the HOD based model. Part of the reason is that the SAM model only has $\sim$50 mocks for testing---these do not fully sample the parameter space as does the HOD model. On the other hand, our model has galaxy number density $\bar{n}$ as a free parameter; the HOD model is more likely to have mocks with low number density dominated by shot noise due to denser sampling. The net effect is to have more outliers for $\xi_{2}$ in terms of model accuracy. We can also notice this, as the SAM model has a more skewed distribution of emulator error. 

With the constructed emulator using both models, we can perform parameter constraint through likelihood analysis using
\begin{equation}
\ln{\mathcal{L}} = -\frac{1}{2}(\xi_{\text{th}}-\xi_{\text{obs}})C^{-1}(\xi_{\text{th}}-\xi_{\text{obs}}),
\end{equation}
where the subscripts 'th' and 'obs' represent the observables from the theoretical prediction (based on the emulator) and observational dataset, respectively. Depending on the purpose, $\xi_{\text{obs}}$ can be CMASS results or measurements from simulations. Following the previous works, we assume two components for the covariance matrix: one is the observed sample variance, and the other is the emulator error (Figure \ref{fig:emulator_error}). The likelihood analysis is carried out using the nested sampling algorithm (\citealt{Skilling_2004}) via the MultiNest package (\citealt{Feroz_2009, Buchner_2014}). In this computation, we choose 1000 live points \footnote{The points in the parameter space originally drawn from the prior volume to iteratively evaluate the likelihood function and sample the posteriors.} to sample the high-dimensional parameter space until the Bayesian evidence reaches a tolerance value of 0.5. The output can include both the final Bayes evidence and posterior of the unknown model parameters.

\section{Results}\label{sec:results}

In this section, we describe the results using both galaxy-halo connection models. Before we present the constraint on the model parameters using clustering measurements, we first examine the (in)consistencies between the models. Given the galaxy mocks generated using different methods, we do not expect this exploration to be comprehensive, but it can be useful to compare parameters and features that both models can describe. This type of analysis has been noticed in previous works such as \cite{Contreras_2013, Pujol_2017, Hadzhiyska_2020, Contreras_2024}, which compare empirical and physical models including HOD, SAM as well as (sub)halo abundance matching (SHAM) and hydrodynamical simulations.

\subsection{Comparison of HOD and SAM}

\subsubsection{Prediction from SAM}

\begin{figure}
\begin{center}
\includegraphics[width=8.5cm]{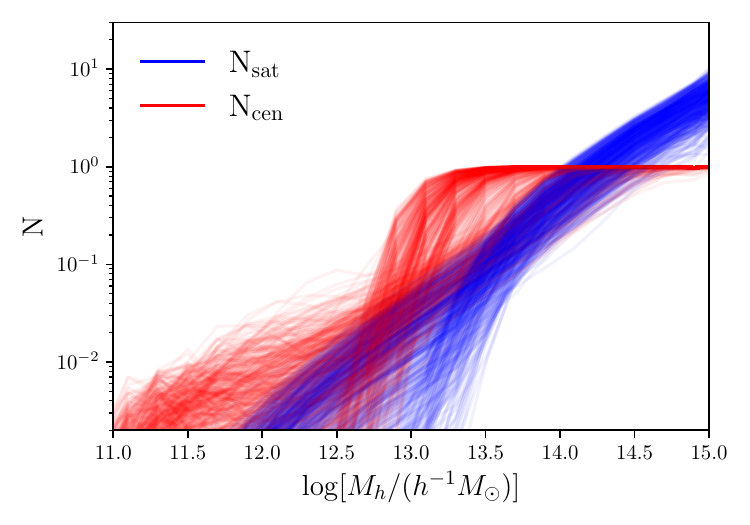}
\caption{The halo occupation distribution predicted by the Galacticus SAM for central and satellite galaxies. The mocks are selected by stellar mass until the sample reaches the required number density, which is allowed to vary in the range [$1.5\times 10^{-4}, 4.5\times10^{-4}$] ($h^{-1}$Mpc)$^{-3}$. }
\label{fig:SAM_true_HOD}
\end{center}
\end{figure}

\begin{figure}
\begin{center}
\includegraphics[width=8.5cm]{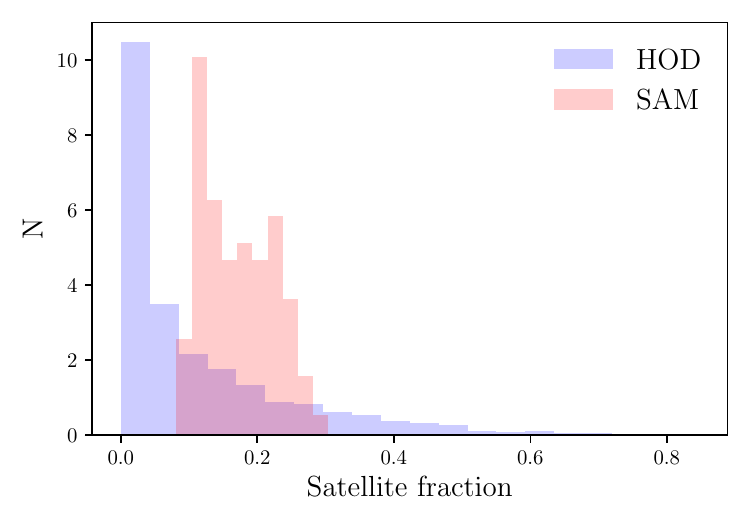}
\caption{Satellite fraction of the galaxy mocks predicted from both HOD and SAM models. The histogram has been normalized by the total number of models.}
\label{fig:fsat_HOD_SAM}
\end{center}
\end{figure}

The GALACTICUS SAM can predict multiple galaxy properties in addition to the phase space information. With the selection based on stellar mass, we show the direct measurement of HOD of these mocks using their host halo information in Figure \ref{fig:SAM_true_HOD}. The result shows both central and satellite galaxies. Since these mocks have a number density in the range [$1.5\times 10^{-4}, 4.5\times10^{-4}$] ($h^{-1}$Mpc)$^{-3}$, the overall amplitude can have a wide distribution. For stellar mass selected galaxies, we can see that most of the mocks have their central occupancy approach unity at the massive end, i.e. the most massive dark matter halos must have a galaxy with high enough stellar mass in the center. Depending on the specific model parameter set, the threshold of halo mass to host a central galaxy has a quite diverse distribution. In addition, it seems to show that the central occupancy has two branches. One has a quick transition from low occupancy to high occupancy, while the other has a slow transition. In terms of the parameterized HOD model, they can correspond to low and high $\sigma_{\log{M}}$. This is consistent with the behavior of the 2PCF in Figure \ref{fig:all_2pcf} that the SAM models have two relatively separate branches. Although not significant, it implies that the sampling with $\sim$500 models in an 18-dimensional parameter space is quite sparse. A brute-force solution by running more models can be a reasonable way to better sample the parameter space and consequently improve the accuracy of the emulator, but it is also worthwhile to investigate other possibilities. On the other hand, the distribution of satellite galaxies shows somewhat more coherent behavior; the occupancy is suppressed at the low mass end at some scale, and the high mass end is close to a power-law function. 

Given the distribution of central and satellite galaxies from the SAM, we present the satellite fraction of these models in Figure \ref{fig:fsat_HOD_SAM}. Similar to the measurement of central occupancy and 2PCF, the satellite fraction of these SAM mocks also shows a weak bimodal distribution, consistent with expectations. By simply splitting galaxy mocks into subsets with high and low satellite fraction, we find that this bimodality is correlated with the dichotomy of the central occupation. Galaxy mocks with higher satellite fraction prefer a smoother transition of central occupation from low to high halo mass and thus a lower clustering amplitude. We then look at the correlation between this behavior and the SAM parameters. We find that only $c_{\text{multi}}$, the multiplicative factor used to scale the cooling rate, shows significant correlation. A higher value of  $c_{\text{multi}}$ increases the overall star formation rate and yields galaxy samples with higher satellite fraction, i.e., satellite galaxies are more likely to be selected with a higher star formation rate than central galaxies. In addition, we note that the sampling of $c_{\text{multi}}$ is done in log space. The distribution of the models is more concentrated at lower values of $c_{\text{multi}}$ and makes the models less distinctive. This can artificially boost the bimodality to some extent, although this does not significantly affect the following analysis. From the figure, we can see that these SAM mocks can cover a range from roughly 8\% to 30\%. On the other hand, the satellite fraction can also be measured from the HOD mocks and we show the results in the same figure. It is clear that the HOD model has a much wider distribution from roughly 0 to 60\%. This makes HOD a flexible model with a larger parameter space for the observational data to be analyzed. On the other hand, it reveals a caveat that the current SAM model may not have enough degrees of freedom for small-scale clustering analysis since parameters like satellite fraction can play a dominant role for one-halo term at this scale.

\subsubsection{Recovering the SAM mock with HOD model}

\begin{figure}
\begin{center}
\includegraphics[width=8.5cm]{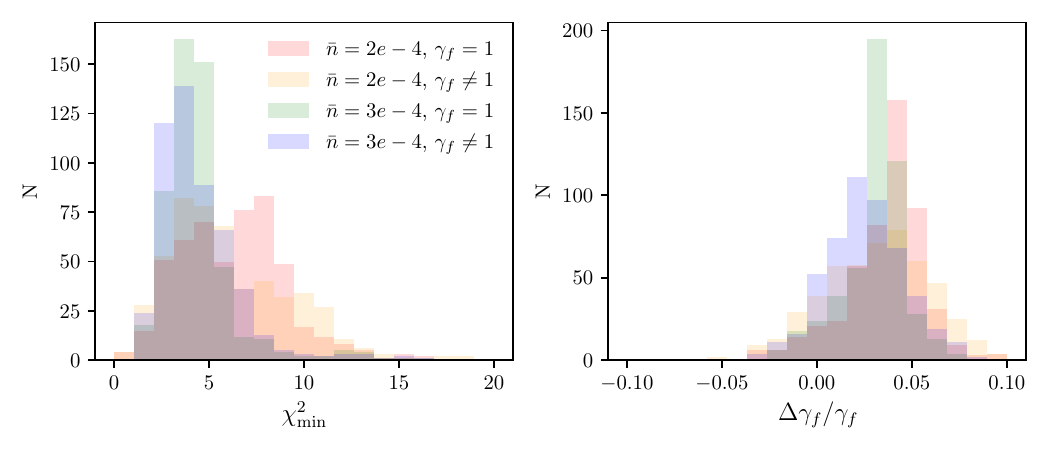}
\includegraphics[width=8.5cm]{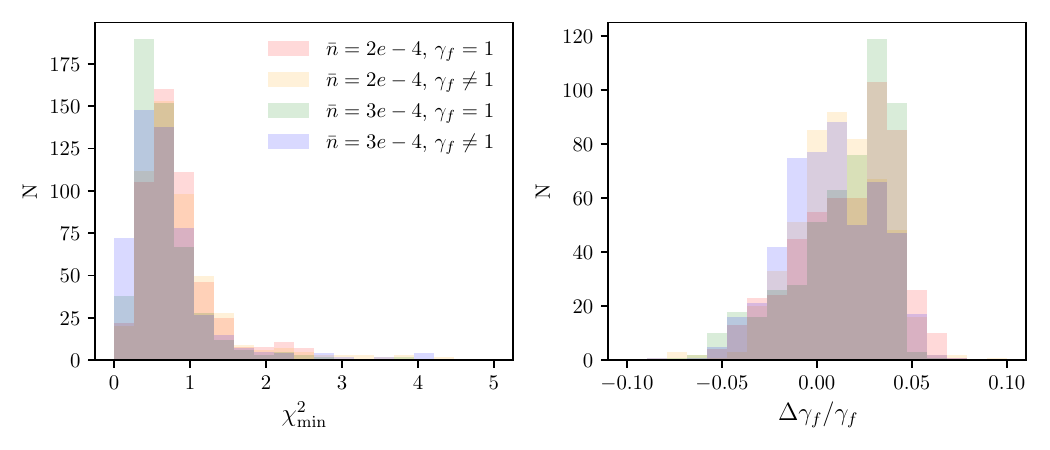}
\caption{The fitting results of all the 500 SAM mocks with the HOD model. $Left:$ distribution of $\chi^{2}_{\mathrm{min}}$ to represent the goodness of fit, $Right:$ distribution of $\gamma_{f}$, the velocity scaling parameter. The top panels show results using 2PCF within scales from 0.1 to 60 $h^{-1}$Mpc, while the bottom panels show results without using the scales below 1$h^{-1}$Mpc.}
\label{fig:HOD_vs_SAM}
\end{center}
\end{figure}

In the previous work (\citealt{Zhai_2023a}), the main goal is to retrieve cosmological information using clustering measurement at non-linear scales. In order to verify the robustness of the result of $f\sigma_{8}$ constraint, which marginalizes over all the nuisance parameters of the empirical HOD model, one can adopt recovery tests on certain galaxy mocks to quantify any possible bias of the parameter constraints by comparing with the input true value. This test has been done using mocks created using subhalo abundance matching (SHAM) from an independent simulation suite, and we find that it verifies that the HOD-based model can yield unbiased cosmological measurement. With the galaxy mocks created by the SAM in this work, we can perform a similar analysis to further examine the robustness of the model. Compared with the previous tests of HOD-based models on SHAM mocks (\citealt{Zhai_2023a}), one critical difference in this work is that both HOD and SAM are applied to the same simulation box of a single cosmology, and thus, we are not able to sample and recover cosmological parameters. In order to proceed, we note that the main cosmological output is the constraint on $f\sigma_{8}$ through the RSD effect. We can assume that this information predominantly comes from the peculiar velocity of the large-scale structure in the universe. In our modeling, the $\gamma_{f}$ parameter can modulate the overall velocity amplitude to effectively make the underlying RSD signal and thus $f\sigma_{8}$ a free parameter, i.e., not fixed at the raw simulation. In this case, our test is to use the HOD-based emulator as our theoretical model, which has $\gamma_{f}$ as a free parameter. Then, we use the prediction from the SAM mock as the test data, which can have the peculiar velocity scaled by different values of $\gamma_{f}$. We perform recovery tests and see if the HOD-based model is able to measure the actual velocity field of the SAM mocks. Note that we can reverse this process to further validate the analysis, i.e., using the SAM-based model to perform recovery tests on the HOD mocks. However, the range of 2PCF from the SAM models is not as wide as HOD models (Figure \ref{fig:all_2pcf}), and some extrapolation is inevitable to derive a biased result. Therefore, we only use the HOD-based model to test the SAM mocks, and it suffices to indicate any possible bias or inconsistency between the models.

In addition to these concerns, we note that in the previous work of testing the HOD model (\citealt{Zhai_2023a}), the SHAM mock was calibrated with a clustering amplitude similar to the observational data. We can choose a similar strategy to test the HOD model using SAM mocks in this work as well by only focusing on a few SAM mocks with a clustering measurement close to CMASS. However, from a theoretical point of view, it is worthwhile to perform a more thorough exploration of the parameter space of the SAM. Therefore we do not just pick a few SAM mocks and run recovery tests, but instead use all of the mocks. Given all these 500 SAM mocks, it is not feasible to run a Monte Carlo Markov Chain or similar statistical inference method on all of them and make sure they all reach sufficient convergence. We decide to focus on the cosmological information or, approximately, the peculiar velocity field. In order to do so, we adopt the likelihood and run an optimization algorithm to find the best-fit parameters instead of the posterior. This is computationally much more economical. For the 2PCF of each SAM mock to be tested, we repeat the optimization process 10 times with different initial positions in the HOD parameter space to ensure that the best-fit parameter is stable. Then, we assume that the parameter set with the lowest $\chi^{2}_{\mathrm{min}}$ of these 10 realizations is the final best-fit parameter. 

In the top panel of Figure \ref{fig:HOD_vs_SAM}, we show the resulting distribution of these best-fit parameters, including $\chi^{2}_{\mathrm{min}}$ and $\gamma_{f}$. We select the SAM mocks by stellar mass with two different number densities $2\times10^{-4}[h^{-1}\mathrm{Mpc}]^{-3}$ and $3\times10^{-4}[h^{-1}\mathrm{Mpc}]^{-3}$, and each case includes two different assumptions for the velocity field: $\gamma_{f}=1$ means the host halos of these SAM galaxies have their velocity field un-scaled, $\gamma_{f}\neq1$ means their halo velocity field is scaled by a factor randomly picked in the range [0.5, 1.5]. From these results, we can see that the value of $\chi^{2}_{\mathrm{min}}$ is quite low, indicating that the best-fit 2PCF of the HOD model is capable of matching the SAM mocks. Although the HOD and SAM models are built on quite different and independent assumptions and procedures, the fitting result quantitatively shows their convergence in terms of galaxy clustering. Given the number of data points in the fit and the number of degrees of freedom, there can be some suspicion that the HOD model is overfitting because many of the models can have a reduced $\chi^{2}$ much lower than 1. However, we should note that the HOD and SAM mocks are not completely independent since they are constructed using the same simulation, i.e., the clustering amplitude can be biased to the same direction due to the sample variance, in this case, the HOD and SAM models just represent different ways of sampling the same halo field. In the likelihood function, the covariance matrix should exclude this contribution and thus lead to a more reasonable $\chi^{2}$ value. But this is beyond the scope of our analysis here. 

$\gamma_{f}$ is the only parameter that both HOD and SAM have in their modeling; the examination of it can have two implications: (1) whether there is any model inconsistency and (2) whether the constraint on the linear growth rate is unbiased. The right-hand side panel of Figure \ref{fig:HOD_vs_SAM} shows the distribution of the fractional offset of the best-fit $\gamma_{f}$. Given the tests of using different number densities and the true input value of $\gamma_{f}$, we can see that there is a 3\% offset in average comparing the inferred value and the true value. Note that the width of the distribution does not show the uncertainty level of these constraints, although they can be correlated. Given the volume of the BOSS galaxies, the number density, and the scales of clustering analysis, the typical uncertainty of the linear growth rate is roughly 5\% (\citealt{Zhai_2019, Zhai_2023a}). This shows that the inconsistency between HOD and SAM is not substantial, and lower than the sample variance. Note that the SAM model is run with hundreds of different parameter sets, indicating that the HOD model can describe the distribution of galaxies created with quite diverse prescriptions of the underlying processes. 

Although the 3\% offset does not appear to be a major inconsistency between the models, we still hope to figure out where it comes from. In order to do so, we rerun the above recovery tests but exclude the clustering measurement at scales below 1$h^{-1}$Mpc, which is dominated by the one-halo terms. We present the results in the bottom panels of Figure \ref{fig:HOD_vs_SAM}. Due to fewer data points, the $\chi^{2}_{\mathrm{max}}$ is lower, consistent with our expectation. On the other hand, the best-fit $\gamma_{f}$ is significantly closer to the truth, i.e., the fractional offset is reduced from 3\% to 1\%. It shows that the modeling at small scales is the dominant source of the offset. We leave more discussion about this in the following section. 

\begin{figure}
\begin{center}
\includegraphics[width=8.5cm]{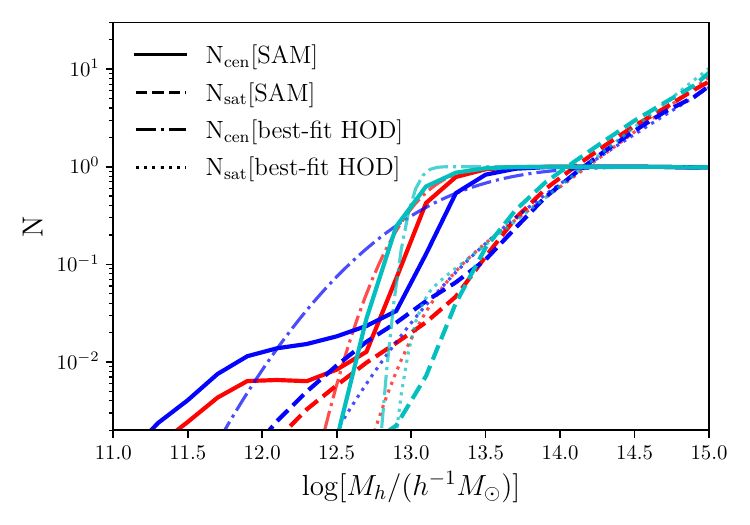}
\caption{Examples of three best-fit HOD models using the parameterization compared with the direct measurement from SHAM mocks, including central and satellite galaxies, respectively.}
\label{fig:SAM_HOD_fit_example}
\end{center}
\end{figure}

In general, we can see that the HOD modeling can provide robust constraints on galaxy mocks in terms of some cosmological measurements. Part of the reason is due to the RSD effect. As long as the information is dominantly from the peculiar velocity field, the way the galaxy population is modeled does not impact the statistical inference after marginalizing over all the nuisance parameters. It also shows that the HOD modeling does not need to accurately and unbiasedly interpret all the properties of the galaxy population. In Figure \ref{fig:SAM_HOD_fit_example}, we show three examples of the HOD measured from the SAM directly, compared to the best-fit value using the parameterization of Eq \ref{eq:NM}. The mismatch of the truth and the best fit is obvious, especially at the low to intermediate halo mass ranges. One of the reasons is due to the HOD parameterization itself. For instance, the central occupancy only allows a smooth transition from 0 to $f_{\mathrm{max}}$. However, the SAM mocks do not necessarily follow this rule even if the galaxies are selected by stellar mass. It can be more severe when the galaxies are selected using parameters other than stellar mass, which can lead to more significant features like bi-model distribution of central occupancy, see for example \cite{Merson_2019, Avila_2020, Zhai_2021} and references therein. Figure \ref{fig:SAM_HOD_fit_example} already shows some hint of bi-modal distribution of our SAM mocks, but the significance is low, and it happens at a low mass scale that the overall galaxy sample is not influenced substantially.

Another caveat in our analysis is in the SAM mock production. We prune branches from the trees below some halo mass scale to accelerate the processing of merger trees, similar to the approach adopted in \cite{Perez_2023}. A direct consequence is a loss of some galaxies in the mock, i.e. the current mock only represents a subsample of galaxies that the model would be able to produce without pruning. It is likely that most of the lost galaxies are less massive than the threshold of our stellar mass-based selection. In Figure \ref{fig:SAM_true_HOD}, we see that the most massive halos still have a galaxy in their center which implies that at least for these central galaxies, the pruning does not have a significant impact. We investigate the influence of this pruning by running a few subsamples of the merger trees with and without pruning. These tests demonstrate that the galaxies that are missed as a result of pruning  occur mainly in the less massive dark matter halos, i.e. the completeness is quite high at the massive end for CMASS-like galaxies. The dark matter halos with a mass above $10^{11}h^{-1}M_{\odot}$ have a completeness of nearly 100\%. Since the pruning is applied at the level of dark matter halos, this completeness is preserved for all the SAM models we have produced. We also identify the missing galaxies from the comparison. As expected, this population is dominated by less massive galaxies. For stellar masses above $10^{10}M_{\odot}$, pruning can result in the loss of around 20\% of galaxies. On the other hand, it is also caused by the fact that pruning can affect the prediction of parameters such as stellar mass, resulting in some underestimation of these masses and affecting the fraction of lost galaxies. For halos of $10^{12}h^{-1}M_{\odot}$, the pruning can underestimate the stellar mass by up to 0.5~dex. However, the overall effect is somewhat mitigated since we select galaxies based on their rank-ordering. This variation is similar to the scatter parameter, $\sigma_{\log{M}}$, in the HOD model. To further investigate the impact, we repeat the above likelihood analysis but randomly remove a fraction (20--30\%) of galaxies from the SAM mock to mimic the effect of pruning. This is similar to our earlier analysis of testing the sample incompleteness for HOD-based analysis in \cite{Zhai_2024}. For a subset of the SAM models, we find some percent-level offset in the recovery test of parameters such as $\gamma_{f}$, but without coherent behavior. This indicates that the pruning of the merger trees does not induce a substantial bias in the cosmological measurements.

\subsubsection{Impact due to sample variance}

\begin{figure}
\begin{center}
\includegraphics[width=8.5cm]{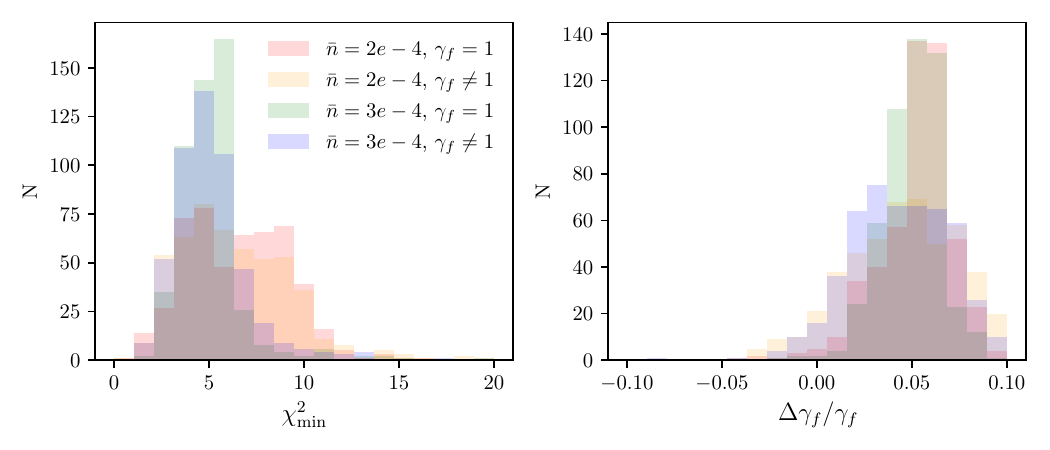}
\caption{The same as the top panel of Figure \ref{fig:HOD_vs_SAM} but the HOD model is built on the fixedAmp\_InvPhase\_001 box, which has the inverted phases relative to the fixedAmp\_001 box. The SAM model for the recovery test is still from the fixedAmp\_001 box.}
\label{fig:HOD_vs_SAM_inverse}
\end{center}
\end{figure}

Our comparison of the HOD and SAM models is based on the same simulation box. The advantage of this is that both models will be affected by the sample variance in precisely the same way, i.e., if the underlying dark matter halos are more/less clustered on some scale and mass range, this signal will propagate to the HOD and SAM model in the same manner. This result can isolate our $\gamma_{f}$ model for velocity scaling in a clean manner, but has the potential of being over-optimistic and underestimating the scatter. In order to further quantify the impact of this sample variance, we can simply repeat our construction of the model and analysis using a different simulation box. To simplify the analysis, we opt to use the fixedAmp\_InvPhase\_001 box from the UNIT suite and build our HOD-based model. Then, we run the new model on the previous SAM mocks, which were constructed using the fixedAmp\_001 box. These two boxes are designed to have the same amplitude of initial power spectrum but inverse phases (\citealt{Chuang_2019}). In terms of clustering measurement, this feature acts to somewhat maximize the difference between boxes, the sample variance. 

In Figure \ref{fig:HOD_vs_SAM_inverse}, we show the recovered results of $\gamma_{f}$. With the same SAM mocks, we do find that the new HOD model has an additional offset of roughly $2\%$ and a net total bias of $5\%$. Given the overall scatter of the distribution, this demonstrates that the sample variance does not play a critical role in the comparison of HOD and SAM models, and the usage of the HOD model for parameter extraction is able to yield robust results.

\subsection{Constraint from BOSS galaxies}

\begin{figure*}
\begin{center}
\includegraphics[width=17.5cm]{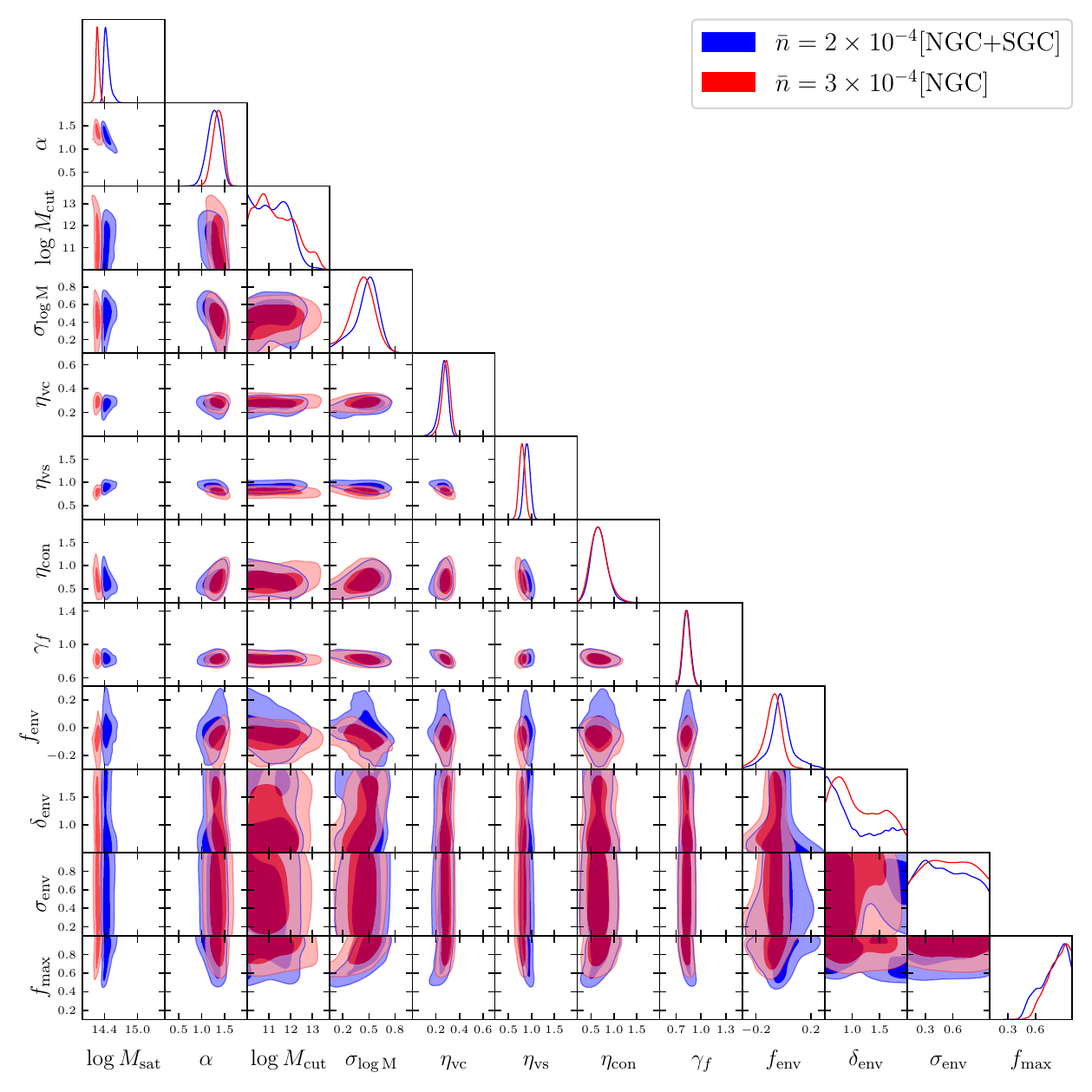}
\caption{1D and 2D constraints on the parameters of the HOD-based model using two subsamples from BOSS-CMASS galaxies. The range of each parameter in the figure has been specified as the prior range, therefore the empty area can indicate the parameter space excluded in the fit.}
\label{fig:constraint_HOD}
\end{center}
\end{figure*}

\begin{figure*}
\begin{center}
\includegraphics[width=17.5cm]{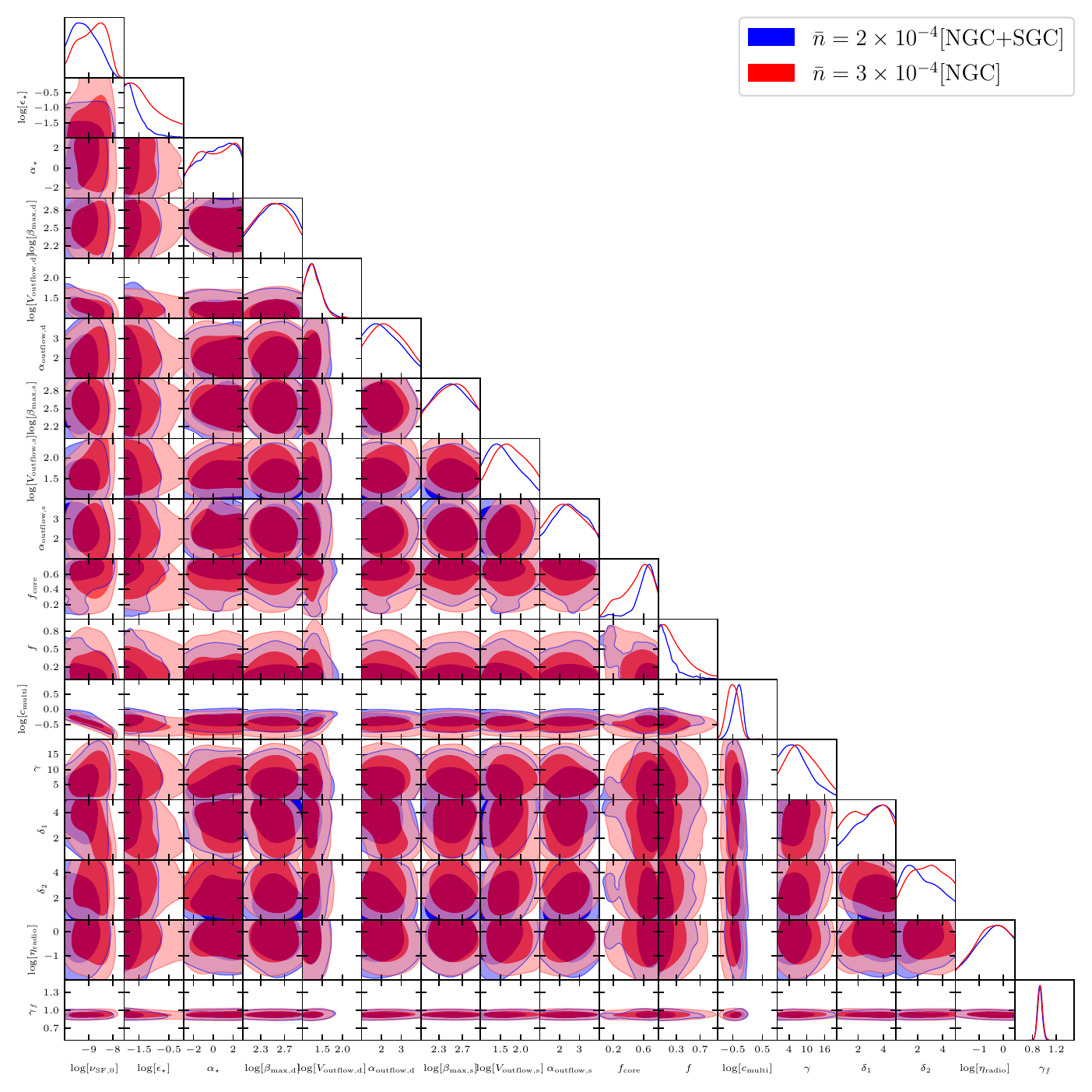}
\caption{The same as Figure \ref{fig:constraint_HOD} but for parameters of the Galacticus SAM.}
\label{fig:constraint_SAM}
\end{center}
\end{figure*}

\begin{figure}
\begin{center}
\includegraphics[width=8cm]{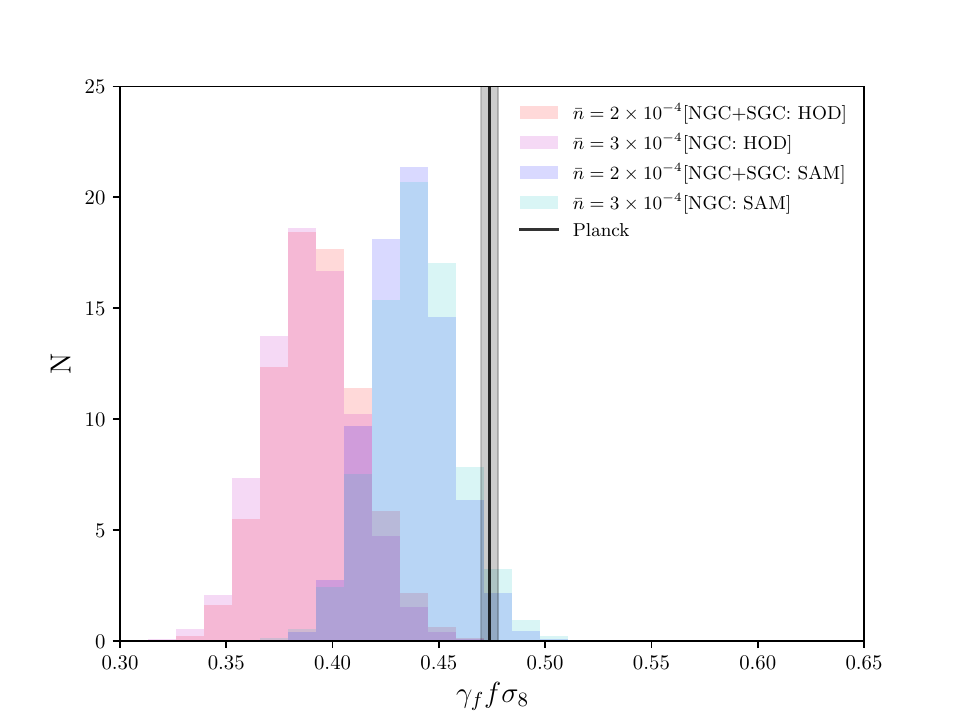}
\caption{Measurement of $f\sigma_{8}$ using both HOD and SAM-based model from BOSS-CMASS galaxies, compared with the Planck prediction at this redshift. Since the model is based on a single UNIT simulation, the measurement is obtained through the parameter $\gamma_{f}$ as explained in the text.}
\label{fig:constraint_fsigma8}
\end{center}
\end{figure}

\begin{figure*}
\begin{center}
\includegraphics[width=17.5cm]{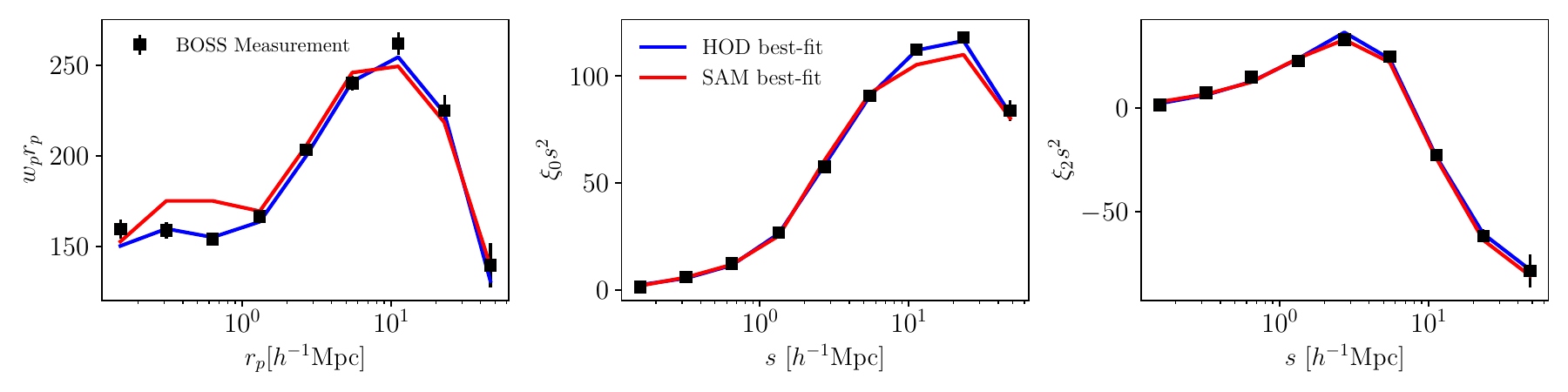}
\caption{Best-fit model of $w_{p}$ (left), $\xi_{0}$ (middle), and $\xi_{2}$ (right) for both HOD and SAM-based models using galaxy sample from the CMASS NGC+SGC with a number density of $2\times10^{-4}$[$h^{-1}$Mpc]$^{-3}$. The value of $\chi^{2}$ for the fit is 21.4 for HOD-based model and 10.9 for SAM-based model. Note that the figure shows only the uncertainty from the data covariance. The above goodness of fit includes contributions from both the data and emulator error. The SAM model has a larger emulator error to dilute the total $\chi^{2}$ and reveals a better fit. But this is not true when we only consider the data covariance.}
\label{fig:2pcf_bestfit}
\end{center}
\end{figure*}

Using 2PCF emulators based on the HOD and SAM mocks, we perform model constraints through clustering measurements from 0.1 to 60 $h^{-1}$Mpc of the BOSS-CMASS galaxies , and present the results in Figure \ref{fig:constraint_HOD} and \ref{fig:constraint_SAM}. We show results using two samples; one is from Aemulus V (\citealt{Zhai_2023a}) where we select galaxies by absolute magnitude to have a constant number density $2\times10^{-4}$[$h^{-1}$Mpc]$^{-3}$ in the redshift range $0.48<z<0.62$ for both north and south Galactic cap (NGC and SGC respectively). The other one is from the analysis of \cite{Zhai_2023b}, where we increase the number density to $3\times10^{-4}$[$h^{-1}$Mpc]$^{-3}$ to have more faint galaxies, but only for NGC. Note that we also have a subsample with a number density of $2\times10^{-4}$[$h^{-1}$Mpc]$^{-3}$ in NGC only, but the measurement and analysis result is quite consistent with the NGC+SGC subsample. Therefore, we do not present its result for simplicity.

The overall result for the HOD parameter constraint is quite similar to previous works (\citealt{Zhai_2023a,Zhai_2023b}), which are based on the Aemulus simulation of varying cosmological parameters. For instance, the typical mass scales of dark matter halos to host a satellite galaxy, roughly $10^{14.44}[h^{-1}M_{\odot}]$ and $10^{14.29}[h^{-1}M_{\odot}]$ for the two subsamples respectively, is slighly lower than the previous work by roughly 0.1 dex, well within $1\sigma$. $f_{\mathrm{max}}$, the parameter for the central occupancy at the massive end, both subsamples show a value close to unity but slightly lower, indicating that only some of these massive halos do not necessarily have a CMASS galaxy in the center. We can also extract constraints on other derived parameters, for instance, the satellite fraction, and we find that both subsamples have roughly 10\% satellites, also consistent with previous work. In general, the clustering measurement is able to provide a tight constraint on most of the parameters in the parameterized model. For the cosmological measurement, the key parameter is $\gamma_{f}$, the velocity scaling parameter. We find that both subsamples have consistent results with a preference of roughly 15\% lower than unity, i.e., lower than the raw prediction from GR. Since the peculiar velocity is directly related to the inference of growth rate, we present the measurement of $f\sigma_{8}$ in Figure \ref{fig:constraint_fsigma8}. Clearly, the inferred value is lower than Planck by $3\sim4\sigma$, similar to the previous results. The constraints from both galaxy subsamples are consistent with each other. 

On the other hand, we show the SAM-based result for $f\sigma_{8}$ in the same figure. There are a few features worth noting. First, the measurement is higher than the HOD-based model, but lower than that of Planck. The discrepancy is reduced to about $2\sigma$. Assuming that the peculiar velocity is directly related to the growth rate inference, the HOD model gives a slightly higher $\gamma_{f}$ on SAM mocks in our previous tests, and it indicates that the SAM mocks can averagely have a stronger amplitude of the peculiar velocity. Therefore, the HOD-based model has to make the value of $\gamma_{f}$ higher to match the SAM prediction. However, at scales of a few $h^{-1}$Mpc, which is assumed to be informative for $f\sigma_{8}$ measurement, we find that the HOD-based model and the SAM mocks can agree at a level of roughly 1\% (Figure \ref{fig:HOD_vs_SAM}). Therefore, the offset observed in the BOSS analysis is likely due to some other reasons, but the significance of this deviation is not high. Second, the SAM-based emulator has a much worse accuracy in the 2PCF prediction than the HOD-based model, therefore the total covariance matrix in the likelihood analysis is indeed dominated by theoretical error instead of the data sample variance. However, the resultant constraint on $f\sigma_{8}$ is comparable to or slightly better than the HOD-based model. Part of the reason is that in terms of galaxy clustering analysis, the components and parameters in the SAM model do not have as much flexibility as the HOD model to describe the possible behaviors of 2PCF. The consequence is that the SAM model is not able to fit the data with a goodness of fit comparable to the HOD model, assuming just sample variance. In other words, the limitation of the model components in the SAM construction artificially underestimates the uncertainty of the measurement. Taking this into account, the offset between HOD and SAM-based models does not seem to be significant. 

In the likelihood analysis, we assume that the theoretical error from the emulator is independent of the data sample variance, and the scale bins are also independent across different summary statistics, i.e., the emulator error contributes only to the diagonal terms of the total covariance matrix. This choice may have less of a significant impact on the HOD-based analysis since the emulator error is not dominant compared with the data, but it can have a substantial impact on the SAM-based analysis. We thus repeat our SAM-based analysis, but assuming that the emulator error has the same correlation of different scales as the data covariance matrix. The results show that the final constraint is similar in terms of both parameter constraints and goodness-of-fit. 

In addition to the cosmological constraint expressed through $\gamma_{f}$, we also examine the inferred constraint on the satellite fraction from the SAM-based model. We find that the SAM prefers a value of roughly 20\% for both two galaxy subsamples, which is much higher than the HOD-based result. Note that the HOD-based result is indeed close to the lower limit that SAM mocks can predict (Figure \ref{fig:fsat_HOD_SAM}). This significant offset can have a distinct behavior in the 2PCF measurement. In Figure \ref{fig:2pcf_bestfit}, we show the best-fit 2PCF of our models compared to the CMASS galaxies. We can see that the SAM-based model does give a much higher $w_{p}$ at small scales, which is the most sensitive to the satellite fraction of a sample. On the other hand, the SAM-based model has a slightly lower amplitude at larger scales for both $w_{p}$ and $\xi_{0}$ compared to the data and the best fit from the HOD-based model. From the perspective of model construction, this may suggest possible improvements that the SAM can take into account in future iterations. For instance, adding some empirical parameters to fine-tune the clustering amplitude may help, but this could be suspicious due to the lack of physical motivations and needs more elaboration. In general, the SAM model is not able to produce a good fit to the 2PCF compared to the HOD-based model. The goodness of fit statistics, such as $\chi^{2}_{\mathrm{min}}$, look acceptable, but they are significantly diluted by the theoretical error. In future work, it will be valuable to investigate potential improvement in the suppression of the theoretical error, i.e., the emulator error. For example, add more correlated summary statistics to enlarge the size of the training sample. We can anticipate that the improvement can lead to more information about the underlying components of the SAM model.

In addition to the cosmological constraint and examination of the galaxy population with the HOD-based model, an important output from the SAM-based model is the constraint on the model parameters, as shown in Figure \ref{fig:constraint_SAM}. Compared with Figure \ref{fig:constraint_HOD}, the clustering measurement has a much weaker constraint on most of the SAM parameters, consistent with our initial expectation since many astrophysical processes do not have a substantial impact on scales of a few $h^{-1}$Mpc and above. This feature is similar to some of the earlier work. For instance, in \cite{vanDaalen_2016}, the authors investigate the Munich SAM with the projected correlation function at low redshift and find that only 5 out of 17 parameters can be constrained using clustering measurement only, and 2 of them are more important than the others, but the combination with abundance information such as stellar mass function can significantly improve the results.  

In our analysis, we find that the clustering measurement from CMASS galaxies can provide a tight constraint on $\epsilon_{\star}$, $V_{\mathrm{outflow, d}}$, $f_{\mathrm{core}}$, $f$ and $c_{\mathrm{multi}}$. The other parameters may have the best-fit value centered in the prior range, but the uncertainty is quite large. Thus, the result failed to exclude a significant amount of parameter space. Our model has different implementations of the astrophysical components compared to the previous works, but the result reveals some similarities. For instance, some parameters governing star formation in the disk and spheroid, like $\epsilon_{\star}$ and $V_{\mathrm{outflow, d}}$, are sensitive to the clustering change. A few parameters in the prescriptions for the gas cooling process can also play a role in galaxy clustering. In addition, the constraint on $c_{\mathrm{multi}}$ seems to show that the data can place a constraint on the overall star formation rate of the galaxies. This parameter also shows some degeneracy with $\nu_{\rm{SF,0}}$, the normalization factor for star formation in the disk.

Throughout the analysis for both HOD and SAM-based analysis, we assume that the number density of the galaxy sample is fixed at the measured value in our model. This is the only information on the abundance of the sample we have used. Given the previous works on similar topics, it is possible that we can get better constraints when we add more data from the galaxy population, such as the stellar mass function or luminosity function. They have been demonstrated to be able to provide a tight constraint on many of the SAM parameters. Combining these measurements with the clustering statistics can potentially yield a constraint on both cosmology and galaxy formation physics with high accuracy. We will leave such investigations in future works.

Compared with the HOD-based model, the SAM-based model has another advantage in that a single parameter set for galaxy formation can produce galaxy mocks for selections of different galaxy types, number densities, and redshifts. For example, our Galacticus mock has parameters including star formation rate, metallically, morphology, luminosity in SDSS $ugriz$ bands, and so on. In the ideal case, it allows for galaxy selections that can mimic exactly any galaxy sample at this redshift. Ultimately, with galaxy samples at different redshifts, this can make the comparison yield a robust model for galaxy formation over the entire redshift range that the observation can cover. In our analysis, the subsamples with different number densities seem to show quite converged constraints on the model parameters. This partially demonstrates the perspective of the SAM approach in modeling both cosmology and galaxy formation.

\section{Discussion and Conclusion}\label{sec:conclusion}

The clustering analysis of galaxies from current cosmological surveys has been demonstrated to be able to extract the underlying information of both the cosmological model and galaxy formation physics. As surveys have improved with larger volumes and better control of the systematics, it is necessary to fully examine the model and minimize the possible bias in the parameter inference. At small scales, this result is highly sensitive to how we understand and characterize the connection between galaxies and their dark matter halos (\citealt{Romeo_2020, Romeo_2023}). Due to simplicity and flexibility, empirical models such as HOD have been applied widely for this purpose to place constraints on the fundamental cosmological parameters and summarize some features of the galaxy population. On the other hand, analyzing the same galaxy sample with an independent modeling approach is of equal importance since it can provide information that traditional HOD does not have. Our work in this paper pushes along this direction by employing the SAM for galaxy formation and evolution to interpret the same observational data as the HOD approach.

The HOD and SAM models aim to solve the same problem---to describe how galaxies are correlated with dark matter halos. The HOD model focuses on the spatial distribution determined by galaxy phase space information, while the SAM is motivated from a more astrophysics-based point of view. This makes the SAM a more physical model to describe the process of galaxy formation in the cosmological background. The cost for this is in the computational demand, which makes it non-trivial to push the application of SAMs to larger scales. The HOD model, on the other hand, can be understood as an approximation of the SAM but with many additional empirical parameters that may lack full physical origins. From this point of view, the SAM is an actual and physical model for the baryonic processes in the modeling of galaxy clustering at different scales, and the multiple components of galaxy formation and evolution can be described in a more self-consistent manner. This makes the empirical models and physical models complementary to each other. Therefore, it is necessary to investigate the convergence of these two independent approaches using the same analysis methodologies.

Our work consists of two separate parts, the first one is a cross-check and validation of the HOD-based and SAM-based model, i.e. the convergence test. We do find some inconsistency between the models, but in terms of cosmological measurement, the bias between them is quite small. The second part is to utilize the clustering measurement, as for HOD analyses, to constrain the SAM parameters and extract the growth rate measurement. In order to better compare with the result from the HOD-based model, the ideal methodology is to apply the SAM on a simulation suite with varying cosmological parameters to build the emulator in the parameter space of both cosmology and SAM parameter, i.e., in the same philosophy as the HOD-based approach. However, such a simulation suite with enough volume, resolution, and merger trees is not available. We have to compromise to focus on a single cosmological simulation and sacrifice the full exploration of the correlation between cosmological parameters and the SAM parameters. Rather, we use the parameterized $\gamma_{f}$ parameter to approximately focus on the measurement of the linear growth rate. On the other hand, it points to the requirement of such simulations for future works, which will be able to provide valuable information to build the correct theory of galaxy formation in the cosmological background.

In the application of the HOD-based and SAM-based models to the CMASS data, we find that they give similar constraints on $f\sigma_{8}$ with some minor offset, both lower than the prediction from the Planck experiment. As two models built upon independent and different modeling assumptions give similar results, this validates the earlier works that the clustering amplitude or the growth rate is lower than expected (\citealt{Leauthaud_2017,Wibking_2020,Lange_2021,Zhai_2023a}). Since more and better data from surveys like DESI (\citealt{DESI_2016}) and Euclid (\citealt{Laureijs_2011}) will be available in the near future, the models developed in this work can be directly applied and provide a more comprehensive study.

In addition, we should note that the uncertainty in the constraint on the SAM-based model may not be fully reasonable. It can be both under and over-estimated. As in Figure \ref{fig:emulator_error}, the emulator error from SAM is much worse than HOD-based model, therefore the total covariance matrix is indeed dominated by the theoretical error instead of the data sample variance, this can artificially boost the error of the final constraint with an acceptable goodness of fit. On the other hand, the SAM does not have many parameters that can directly affect clustering as the HOD-based model. In other words, the SAM is much less flexible and can indeed fail to fit the data, as we can see from the best-fit model in Figure \ref{fig:2pcf_bestfit}. The consequence is to underestimate the uncertainty. For example, one typical parameter in the clustering analysis is the satellite fraction. For the HOD-based model, as long as we can assume that the host halo is complete down to some mass scale, the number of satellite galaxies controlled by the occupation parameters can be arbitrarily high or low, or equivalently, the mass resolution for satellites is infinitely high. It is thus easy for a HOD model to fit the 1-halo terms due to this flexibility. On the other hand, the satellites in the SAM model follow the evolution of the halos and subhalos, and are directly affected by the mass resolution limit. This can restrict the range of the 1-halo terms that the SAM can model. Given this performance, one way to improve the performance of the SAM is to introduce additional phenomenological parameters as the HOD model, either during the production of the mock or in the post-processing, just as we implement the velocity scaling parameter in this work. In terms of data fitting of clustering measurement, it will be able to yield improved results, but the cost is to lose predictive power as a physical model for galaxy formation and evolution.

Our current emulator of the SAM 2PCF has a much worse accuracy compared with the HOD-based model. One reason is the sparse sampling in the parameter space, thus a possible solution is to add more training models, i.e. simply double the size of the SAM mocks. However, our test using the HOD-based model shows that it may needs a few thousands of training sample to reach a percent level of accuracy. Since our SAM model has more parameters, the final size of the required training sample can be even larger. Given the current performance of the code, the overall computational cost makes this brute-force solution unfeasible. On the other hand, our analysis result using BOSS galaxies shows that the area in the current parameter space for training sample, high $v_{\mathrm{SF,0}}$ and high $c_{\mathrm{multi}}$ is not favored. This implies that adding more training mocks may not be the ideal solution. One way worth investigating is to adopt more statistics and their correlations, for instance consider beyond standard 2PCF, add correlations between galaxy mocks and the dark matter halo field, or construct the model at the field level using machine learning algorithms. This may have the power to improve the overall accuracy of the SAM-based model. Another approach is to simply fix a few parameters that are not sensitive to clustering measurements and thus reduce the dimensionality of the parameter space. This may increase the interpolation accuracy and lower the emulator error. However, we find that this still requires a few hundred models for either the HOD or the SAM algorithm, even for only 5 or 6 parameters, to ensure the emulator error is subdominant compared with the data covariance. From this point of view, our work can serve as a pilot investigation to search the parameter space and point to the direction for future work.

In addition to the cosmological parameters, our clustering analysis provides constraints on some of our Galacticus SAM parameters. This is promising since most of the astrophysical processes are assumed to only affect galaxy distribution at smaller scales. However, the result shows that some components, such as star formation in the disk and spheroid, outflow due to stellar feedback, and so on, can influence the clustering amplitude at larger scales. On the other hand, many of these parameters are sensitive to the abundance information. However, in our work, only the galaxy number density is used in the definition of the sample. A clear way to improve this situation would be to apply additional datasets, such as the stellar mass function or luminosity function, as done in many previous works, to calibrate the SAM model. The caveat in our model is that the approximations we use, such as pruning the merger trees of some dark matter halos, can result in an underestimate of some quantities, such as stellar mass. This may not change the rank ordering for our sample selection significantly, but can impact the overall shape and amplitude of statistics such as the stellar mass function. With an improved model in both accuracy and speed, we can anticipate another leap in the constraint on the SAM parameters when both abundance and clustering information are used.

Similar to other SAMs, our Galacticus model is able to provide a variety of galaxy parameters, such as stellar mass we have used, star formation rate, spin, metallicity, and many others. The sampling in the model parameter space can enable the investigation of the related statistics on the model parameters. For instance, how central and satellite galaxies are correlated, how the clustering of galaxies can depend on these parameters. We will present such investigations in separate works.

% End of mnras_template.tex

\section*{Acknowledgements}
ZZ appreciates helpful comments and discussions with Jeremy Tinker and Zheng Zheng. ZZ is supported by NSFC (12373003), the National Key\&D Program of China (2023YFA1605600), and acknowledges the generous sponsorship from Yangyang Development Fund. This work is also supported by the China Manned Space Program
with grant no. CMS-CSST-2025-A04. AB and YW gratefully acknowledge funding from NASA Grant  \#80NSSC24M0021, "Project Infrastructure for the Roman Galaxy Redshift Survey".

This research was enabled in part by support provided by Compute Ontario (computeontario.ca) and the Digital Research Alliance of Canada (alliancecan.ca).

\rm{Software:} Python,
Matplotlib \citep{matplotlib},
NumPy \citep{numpy},
SciPy \citep{scipy}.

\section*{Data Availability}

The galaxy mocks and data are available upon reasonable request to the author.

\appendix

\bibliographystyle{mnras}
\bibliography{emu_gc_bib,software}

\bsp	% typesetting comment
\label{lastpage}
\end{document}